\documentclass[conference]{IEEEtran}
\usepackage{amsfonts}

\IEEEoverridecommandlockouts

\ifCLASSINFOpdf
\else
\fi
\everymath{\small}
\everydisplay{\small}
\usepackage{lipsum}
\usepackage{mathrsfs}
\usepackage{amssymb}
\usepackage{pdfpages}
\usepackage{subfigure}
\usepackage{epsfig}
\usepackage{graphicx}
\usepackage{psfig}
\usepackage{epsf}
\usepackage[cmex10]{amsmath}
\usepackage{booktabs}
\usepackage{fancyhdr}
\usepackage[ruled,vlined,linesnumbered]{algorithm2e}
\usepackage{amsmath}
\usepackage{bm}
\usepackage{setspace}
\usepackage{caption2}   

\begin{document}
\title{Joint Resource Optimization Over Licensed and Unlicensed Spectrum in Spectrum Sharing UAV Networks Against Jamming Attacks}
\author{Rui Ding$^{ \S}$, Fuhui Zhou$^{ \S,^\ddagger}$, Yuhang Wu$^{ \S}$, Qihui Wu$^{ \S}$, and Tony Q. S. Quek$^\star$\\
$^{\S}$Nanjing University of Aeronautics and Astronautics, China,\\
$^\star$Singapore University of Technology and Design, Singapore
\\
Email:
\emph{rui\_ding@nuaa.edu.cn, zhoufuhui@ieee.org, may\_wyh@nuaa.edu.cn}, \\
\emph{wuqihui2014@sina.com, and tonyquek@sutd.edu.sg}

\thanks{
This work was supported in part by the Jiangsu Province Frontier Leading Technology Basic Research Project under Grant BK20222013;
in part by the National Natural Science Foundation of China under Grant 62222107;
 in part by the National Key R\&D Program of China under Grant 2023YFB2904500;
 in part by the Basic Reserch Projects of Stabilizing Support for Specialty Disciplines under Grant No.ILF240041A24;
 in part by the Postgraduate Research \& Practice Innovation Program of Jiangsu Province.
 (\emph{The corresponding author is Fuhui Zhou.})
}

}

\maketitle
\begin{abstract}
Unmanned aerial vehicle (UAV) communication is of crucial importance in realizing heterogeneous practical wireless
application scenarios.
However, the densely populated users and diverse services with high data rate demands has triggered an increasing scarcity of UAV
spectrum utilization.
To tackle this problem, it is promising to incorporate the underutilized unlicensed spectrum with the licensed spectrum to boost network capacity.
However, the openness of unlicensed spectrum makes UAVs susceptible to security threats from potential jammers.
Therefore, a spectrum sharing UAV network coexisting with licensed cellular network and unlicensed Wi-Fi network is considered with the anti-jamming technique in this paper.
The sum rate maximization of the secondary network is studied by jointly optimizing the transmit power, subchannel allocation, and UAV trajectory.
We first decompose the challenging non-convex problem into two subproblems, 1) the
joint power and subchannel allocation and 2) UAV trajectory design subproblems.
A low-complexity iterative algorithm is proposed in a alternating optimization manner over these two subproblems to solve the formulated problem.
Specifically, the Lagrange dual decomposition is exploited to jointly optimize the transmit power and subchannel allocation iteratively.
Then, an efficient iterative algorithm capitalizing on successive convex approximation is designed to get a suboptimal solution for UAV trajectory.
Simulation results demonstrate that our proposed algorithm can significantly improve the sum transmission rate compared with the benchmark schemes.
\end{abstract}
\begin{IEEEkeywords}
Spectrum sharing, unlicensed spectrum, spectrum allocation, trajectory optimization.
\end{IEEEkeywords}
\IEEEpeerreviewmaketitle

\section{Introduction}
\IEEEPARstart
Due to the high mobility, flexible deployment, and low cost, unmanned aerial vehicles (UAVs) have
played important roles in realizing heterogeneous practical wireless application scenarios, such as surveillance and monitoring, temporary base station, aerial imaging, cargo delivery, etc \cite{xxxx}.
However, with the forthcoming  sixth-generation (6G) era,
densely populated users exhibit a significant demand for broadband wireless communications, and network operators are expected
to support diverse services with high wireless data demands such as multimedia streaming and video downloads  \cite{marco}.
Consequently, this has led to an increasing scarcity of UAV spectrum utilization.

Spectrum sharing is a promising technique to solve this problem.
However, most existing UAV networks need to share spectrum with licensed operators to enhance the transmission capacity since UAVs have limited dedicated spectrum available.
Despite this, the incredible increase in connected appliances and downloaded applications has pushed mobile operators to the limits of the licensed spectrum \cite{6gnru}, \cite{6gnru2}. Moreover, the costly and scarce licensed spectrum pose significant challenges for operators to allocate new spectrum specifically for UAV communications.
These challenges have triggered the exploration of the underutilized unlicensed spectrum to extend the available spectrum resources.


Recent advances in unlicensed wireless fidelity (Wi-Fi) spectrum such as Wi-Fi 7 has focused on keeping up with the increasing data rate demand \cite{wifi7}.
In particular, the introduction of OFDMA and wider channel bandwidth in the 6 GHz band can provide higher user service density and reduce latency \cite{wxb}.
 Moreover, the multi-link operation allows for seamless access to multiple channels with a theoretical data rate of 30 Gbps \cite{6gnru}.
Therefore, evolving to leverage the underutilized unlicensed spectrum through spectrum sharing is promising to expand available spectrum resources for UAV communications.

Inspired by the aforementioned potential benefits of unlicensed Wi-Fi spectrum, it is expected that unlicensed spectrum can be utilized to further enhance the system performance and user connectivity in spectrum-sharing networks.
For instance, the authors in \cite{shermen} studied the uplink sum-rate maximization problem in a multi-cell UAV-cellular network, where the unlicensed spectrum was employed by UAV-BSs to increase the achievable rate.
However, the resource allocation schemes proposed in \cite{shermen} were unsuitable for UAV air-ground spectrum sharing since the transmission channel in UAV-cellular network is different from air-ground line-of-sight (LoS) link.
Moreover, a UAV-assisted Internet of vehicles system coexists with a Wi-Fi system was investigated in \cite{iot_magz} to maximize the user satisfaction.
However, this study was limited to the case of Internet of vehicles and cannot be applied to general cases with different network topologies.
Moreover, the interference to licensed network was assumed negligible and the transmit power was considered fixed, which is generally over optimistic.
Furthermore, UAVs may suffer from security threats from jammers due to the open characteristic of the unlicensed spectrum \cite{ding}.
Therefore, to address the aforementioned issues and reap the advantages of unlicensed spectrum through spectrum sharing to enhance the data rate and alleviate the spectrum scarcity, it is important to investigate the spectrally-efficient resource allocation over licensed and unlicensed spectrum in spectrum sharing UAV networks.

In this paper, a spectrum sharing UAV network against jamming attacks is investigated.
We formulated a sum rate maximization problem taking into account the sharing of licensed and unlicensed spectrum with the existence of a potential jamming to jointly optimize the spectrum allocation, power allocation, and UAV trajectory.
An efficiently iterative algorithm is proposed to tackle the formulated intractable non-convex optimization
problem.
Simulation results reveal that our proposed algorithm can drastically enhance the sum transmission rate compared with the benchmark
schemes.

The remainder of this paper is organized as follows.
In Section II, the system model is presented.
The problem formulation is shown in Section III.
 Section IV presents the proposed iterative algorithm to jointly optimize the transmit power, subchannel allocation, and UAV trajectory.
The simulation results are shown in Section V.
Finally, this paper is concluded in Section VI.

\section{System Model}
\subsection{Scenario Description}
\begin{figure}
\centering
\includegraphics[width=3.343in]{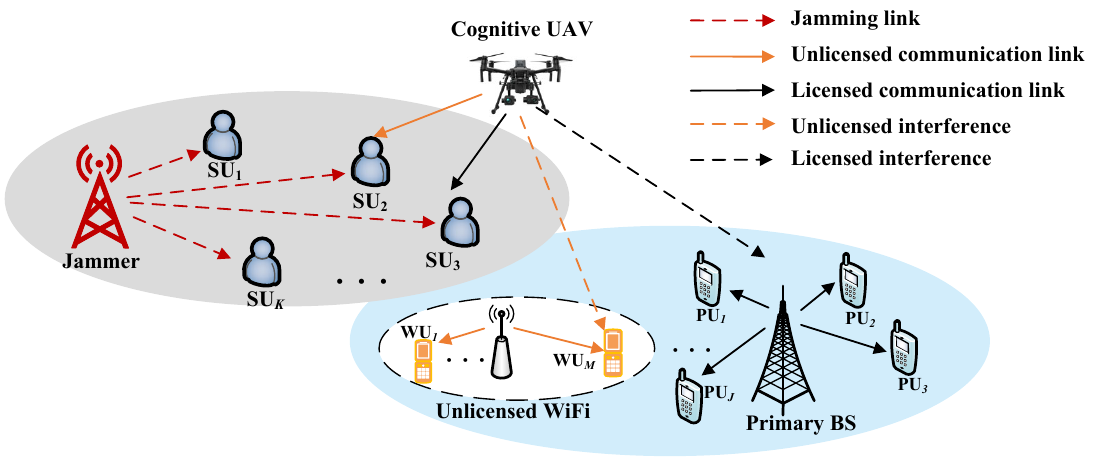}
\caption{The spectrum sharing UAV network with dynamic user requirements.} \label{fig.1}
\end{figure}

As shown in Fig. \ref{fig.1}, a anti-jamming spectrum sharing UAV network is considered, which consists of a licensed cellular network, an unlicensed Wi-Fi network, and a cognitive UAV network.
Specifically, the primary network comprises one primary base station (PBS) and $J$ primary users (PUs). The Wi-Fi network contains one Wi-Fi access point and $M$ Wi-Fi users (WUs).
One cognitive UAV is considered as the secondary base station to serve the ground secondary users (SUs).
Let $k\in \mathcal{K}\triangleq \{1,2,...,K\}$, $m\in \mathcal{M}\triangleq \{1,2,...,M\}$, and $j\in \mathcal{J}\triangleq \{1,2,...,J\}$
denote the set of SUs, WUs, and PUs, respectively.
The jammer equipped with $N_{J}$ antennas is located near the SUs, attempting to send faked or replay jamming signals to degrade legitimate communication performance.
In order to serve more users and provide better quality of services, a wideband spectrum is divided into an orthogonal set of finite licensed subchannels with uniform bandwidth.
For reliable signal transmission from C-UAV to SUs, each SU is only allowed to access one licensed subchannel, and each licensed subchannel is assigned to at most one SU.
Moreover, the SUs can reinforce the data rate through operating in the unlicensed Wi-Fi spectrum in order to support the minimum transmission data rate of the $k$th SU $R_{k}^{\mathrm{min}}$.
The bandwidth $B_{u}$ of the unlicensed channel is divided by the cognitive UAV network into a set of finite subchannels with uniform bandwidth for efficient resource management \cite{shermen}.

A three-dimensional Cartesian coordinate system is considered.
The horizontal positions of the PBS, the $j$th PU, the $k$th SU, and the C-UAV are denoted as
${\mathbf{w}_{b}}=({{x}_{b}},{{y}_{b}})$, ${\mathbf{w}_{p,j}}=({{x}_{p,j}},{{y}_{p,j}})$,
 ${\mathbf{w}_{s,k}}=({{x}_{s,k}},{{y}_{s,k}})$ and ${\mathbf{q}_{c}}=({{x}_{c}},{{y}_{c}})$, respectively.
Without loss of generality, it is assumed that the UAVs fly at a constant vertical height $H_{u}$.
The total transmission time interval is within a duration of $T$, and $T$ is divided into $N$ equal-length time intervals,
where each time interval is given by ${{\delta }_{t}}=\frac{T}{N}$.
The status of the UAVs can be regarded as static since $\delta_{t}$ is sufficiently small \cite{ding}. Let $n\in \mathcal{N}\triangleq \{1,2,...,N\}$ denote the set of time steps.
The dynamic position of the C-UAV can be formulated as
\begin{subequations}
\begin{align}
&x_{c}[n+1]=x_{c}[n]+{{v}_{c}[n]}\cos (\phi_{c} [n]),\\
&y_{c}[n+1]=y_{c}[n]+{{v}_{c}[n]}\sin (\phi_{c} [n]),
\end{align}
\end{subequations}
where $\phi_{c}[n]$ and $v_{c}[n]$ represent the direction and the flying speed of the C-UAV at time step $n$, respectively.


\subsection{Data Transmission in Licensed Spectrum }
The information symbol for the $k$th SU transmitted by the C-UAV is represented by $x_{k}^{s}$ and $\mathbb{E}[\left|x_k^s\right|^2]=1$.
The received signal of the $k$th SU in the licensed spectrum consists of the signal from the C-UAV, the interference from the PBS, and the jamming signal from the jammer, which is given as
\begin{align}
y_{k,j}^{\mathrm{lic}}[n]=&h_{k}^{\mathrm{us}}[n]\sqrt{p_{k,j}^{\mathrm{lic}}[n]}x_{k}^{s}[n] \nonumber+h_{k}^{\mathrm{ps}}[n]\sqrt{p_{j}^{\mathrm{P}}[n]}{{s}_{j}}[n] \nonumber \\
&+{{\mathbf{g}}_{k}^H}[n]{{\mathbf{w}}_{k}}[n]x_{k}^{\mathrm{jam}}[n]+n_{k}^{{}},
\end{align}
where $p_{k,j}^{\mathrm{lic}}[n]$ denotes the transmit power from the C-UAV to the $k$th SU at the $j$th licensed subchannel.
${p_{j}^{\mathrm{P}}[n]}$ represents the transmit power from the PBS to the $j$th PU and $s_{j}[n]$ denotes the normalized signal symbol for the $j$th PU.
The channel between the C-UAV and the $k$th SU, and between the PBS and the $k$th SU are denoted by $h_k^{\mathrm{us}}[n]$ and
$h_{k}^{\mathrm{ps}}[n]$, respectively.
In particular, the wireless channel between the UAV and the ground users is dominated by the LoS link.
Both the distance-dependent path loss with exponent $\phi>2$ and small-scale Rayleigh fading are considered for the channel model \cite{yuhang}.
Moreover, a multi-antenna jammer attempts to interrupt the communications by sending the jamming signal ${\mathbf{w}}_{k}[n]x_k^{\mathrm{jam}}[n]$ to the legitimate user, where $\mathbb{E}[|x_{k}^{\mathrm{jam}}|^2]=1$ represents the normalized
signal symbol and $\mathbf{w}_{k}[n] \in \mathbb{C}^{N_J \times 1}$ is the beamforming vector of the jammer.
The channel between the jammer and the $k$th SU is denoted by ${{\mathbf{g}}_{k}}[n] \in \mathbb{C}^{N_J \times 1}$.
$n_k \sim \mathcal{C} \mathcal{N}\left(0, \sigma_k^2\right)$ is the additive white Gaussian noises (AWGNs) at the $k$th SU.

Then, we have the SINR of the $k$th SU on the $j$th licensed subchannel expressed as
\begin{align}
\mathrm{SINR}_{k,j}^{\mathrm{lic}}[n]=\frac{|h_{k}^{\mathrm{us}}[n]{{|}^{2}}p_{k,j}^{\mathrm{lic}}[n]}{|h_{k}^{\mathrm{ps}}[n]{{|}^{2}}p_{j}^{\mathrm{P}}[n]+|{{\mathbf{g}}_{k}^H}[n]{{\mathbf{w}}_{k}}[n]{{|}^{2}}+{{\sigma }_k^{2}}}.
\end{align}
Accordingly, the data rate of SU $k$ is given as
\begin{align}
R_{k,j}^{{\mathrm{lic}}}[n]={{{\rho }_{k,j}^{\mathrm{lic}}}[n]{{\log }_{2}}(1+{\mathrm{SINR}}_{k,j}^{{\mathrm{lic}}}[n])},
\end{align}
where the binary variable ${\rho }_{k,j}^{\mathrm{lic}}[n]$ is adopted to characterize the licensed spectrum allocation strategy of SUs.
Specifically, the $j$th licensed subchannel is used by the $k$th SU at time step $n$ when ${\rho }_{k,j}^{\mathrm{lic}}[n]=1$, otherwise, ${\rho }_{k,j}^{\mathrm{lic}}[n]=0$.

\subsection{Data Transmission in Unlicensed Spectrum}
Similarly, the received signal of the $k$th SU on the unlicensed spectrum can be given as
\begin{align}
y_{k,m}^{\mathrm{unlic}}[n]=&h_{k}^{\mathrm{us}}[n]\sqrt{p_{k,m}^{\mathrm{unlic}}[n]}x_{k}^{s}[n] \nonumber+h_{k}^{\mathrm{ws}}[n]\sqrt{p_{m}^{\mathrm{Wifi}}[n]}s_{m}^{\mathrm{Wifi}}[n] \nonumber \\
&+{{\mathbf{g}}_{k}}[n]{{\mathbf{w}}_{k}}[n]x_{k}^{\mathrm{jam}}[n]+n_{k}^{{}},
\end{align}
where $p_{k,m}^{\mathrm{unlic}}[n]$ denotes the transmit power from the C-UAV to the $k$th SU on the $m$th unlicensed subchannel.
${p_{m}^{\mathrm{Wifi}}[n]}$ represents the transmit power from the Wi-Fi to the $m$th WU.
$s_{m}^{\mathrm{Wifi}}[n]$ denotes the normalized information symbol for the $m$th WU transmitted by the Wi-Fi access point.

Then, we have the SINR of the $k$th SU on the unlicensed spectrum expressed as
\begin{small}
\begin{align}
\mathrm{SINR}_{k,j}^{\mathrm{unlic}}[n]=\frac{|h_{k}^{\mathrm{us}}[n]{{|}^{2}}p_{k,j}^{\mathrm{unlic}}[n]}{|h_{k}^{\mathrm{ws}}[n]{{|}^{2}}p_{m}^{\mathrm{Wifi}}[n]+|{{\mathbf{g}}_{k}^H}[n]{{\mathbf{w}}_{k}}[n]{{|}^{2}}+{{\sigma }_k^{2}}}.
\end{align}
\end{small}
Accordingly, the data rate of SU $k$ is given as
\begin{align}
R_{k,m}^{{\mathrm{unlic}}}[n]={{{\rho }_{k,m}^{\mathrm{unlic}}}[n]{{\log }_{2}}(1+{\mathrm{SINR}}_{k,m}^{{\mathrm{unlic}}}[n])},
\end{align}
where the binary variable ${\rho }_{k,m}^{\mathrm{unlic}}[n]$ is adopted to characterize the unlicensed spectrum allocation strategy of SUs.
Specifically, the $m$th unlicensed subchannel is used by the $k$th SU at time step $n$ when ${\rho }_{k,m}^{\mathrm{unlic}}[n]=1$, otherwise, ${\rho }_{k,m}^{\mathrm{unlic}}[n]=0$.

A minimum data transmission rate $R_k^{\mathrm{min}}$ is required by each SU for its application.
The C-UAV allows to access resources from the unlicensed spectrum to enhance SU data rate when $R_{k}^{\mathrm{lic}}[n] < R_{k}^{\mathrm{min}}$.
Therefore, the minimum transmission rate for the $k$th SU is achieved through the constraint $1/N\sum\limits_n^{} {{R_k}[n]}  \ge R_k^{\min }$, where $R_k[n] = \sum\limits_j^{}R_{k,j}^{\mathrm{lic}}[n]+\sum\limits_m^{}R_{k,m}^{\mathrm{unlic}}[n]$.

\section{Problem Formulation}\label{section3}
In this paper, in order to efficiently utilize the spectrum resource, protect the users from harmful interference, and guarantee the transmission rate requirements of SUs, a joint licensed and unlicensed spectrum allocation, power optimization, and UAV trajectory optimization problem is formulated

\begin{small}
\allowdisplaybreaks
\begin{subequations}
\begin{align}
   \textbf{P}_{1}: &\underset{\mathcal{A},\mathcal{B},\mathcal{P},\mathcal{U},\mathcal{Q}}{\mathop{\max }}\,\frac{1}{N}\sum\limits_{n=1}^{N}{\sum\limits_{k=1}^{K}{{{R}_{k}^{}}[n]}} \\
  \mathrm{s.t.} \,  &\mathrm{C1}: \frac{1}{N} \sum_n R_k[n] \geq R_k^{\min }, \forall k, \\
 &\mathrm{C2}: \frac{1}{N} \sum_n \sum_k \rho_{k, j}[n]\left|h_j^{{\mathrm{up}}}[n]\right|^2 p_{k, j}^{\text {lic}}[n]<\Gamma_j^{{\mathrm{lic}}}, \forall j, \\
 &\mathrm{C3}: \frac{1}{N} \sum_n \sum_k \rho_{k, m}[n]\left|h_m^{\mathrm{uw}}[n]\right|^2 p_{k, m}^{\mathrm{unlic}}[n]<\Gamma_m^{ {\mathrm{unlic}}}, \forall m, \\
& \mathrm{C4}: \sum_k \rho_{k, j}^{{\mathrm{lic}}}[n] \leq 1, \forall j, \forall n, \\
& \mathrm{C5}: \sum_j \rho_{k, j}^{ {\mathrm{lic}}}[n] \leq 1, \forall k, \forall n, \\
& \mathrm{C6}: \sum_k \rho_{k, m}^{ {\mathrm{unlic}}}[n] \leq 1, \forall m, \forall n, \\
& \mathrm{C7}: \rho_{k, {\mathrm{lic}}}^{ {\mathrm{lic}}}[n] \in\{0,1\}, \forall k, \forall j, \forall n, \\
& \mathrm{C8}: \rho_{k, m}^{ {\mathrm{unlic}}}[n] \in\{0,1\}, \forall k, \forall m, \forall n,\\
& \mathrm{C9}: \sum_k \sum_j \rho_{k, j}^{{\mathrm{lic}}}[n] p_{k, j}^{{\mathrm{lic}}}[n] \leq P_{\max }^{{\mathrm{lic}}}, \forall n, \\
& \mathrm{C10}: \sum_k \sum_m \rho_{k, m}^{{\mathrm{unlic}}}[n] p_{k, m}^{{\mathrm{unlic}}}[n] \leq P_{\max }^{{\mathrm{unlic}}}, \forall n, \\
& \mathrm{C11}: \|\mathbf{q}[n]-\mathbf{q}[n-1]\|^2 \leq\left(V_{\max } \delta_t\right)^2, \forall n,
\end{align}
\end{subequations}
\end{small}
where the licensed and unlicensed subchannel allocation variable set as $\mathcal{A}=\{{{\rho }_{k,j}^{\mathrm{lic}}}[n],\forall k \in \mathcal{K},\forall j \in \mathcal{J},{\forall n \in \mathcal{N}}\}$ and
$\mathcal{B}=\{\rho _{k,m}^{\text{unlic}}[n],\forall k\in \mathcal{K},\forall m\in \mathcal{M},\forall n\in \mathcal{N}\}$, respectively.
The licensed and unlicensed transmit power variable set as $\mathcal{P}=\{p_{k,j}^{\mathrm{lic}}[n], \forall k \in \mathcal{K}, \forall j \in \mathcal{J}, \forall n \in \mathcal{N}\}$ and
$\mathcal{U}=\{p_{k,m}^{\mathrm{unlic}}[n], \forall k \in \mathcal{K}, \forall m \in \mathcal{M}, \forall n \in \mathcal{N}\}$, respectively.
The UAV location variable set as $\mathcal{Q} = \{\mathbf{q}[n], \forall n \in \mathcal{N}\}$.

Note that the minimum transmission rate requirement for SUs is achieved through constraint $\mathrm{C1}$.
$\Gamma_j^{ {\mathrm{lic}}}$ and $\Gamma_m^{ {\mathrm{unlic}}}$ in $\mathrm{C2}$ and $\mathrm{C3}$ are the maximum tolerable interference for licensed and unlicensed spectrum, respectively.
$\mathrm{C4}$, $\mathrm{C5}$, and $\mathrm{C7}$ are licensed spectrum allocation constraint such that each SU can share the licensed spectrum with at most one PU and each licensed subchannel can be allocated to at most one SU.
$\mathrm{C6}$ and $\mathrm{C8}$ are unlicensed spectrum allocation constraint such that each unlicensed subchannel can be assigned to at most one SU to avoid multiple access interference.
$P_{\mathrm{max}}^{\mathrm{lic}}$ and $P_{\mathrm{max}}^{\mathrm{unlic}}$ in $\mathrm{C9}$ and $\mathrm{C10}$ are the peak transmit power on the licensed and unlicensed spectrum, respectively.
$\mathrm{C11}$ is the maximum allowable flying speed of the C-UAV.

\section{Proposed Alternative Algorithm for Joint Resource Allocation and Trajectory Optimization}\label{section3}
The formulated problem $\textbf{P}_1$ is non-convex, which generally cannot be solved efficiently by conventional convex
optimization methods.
To facilitate a low computational complexity design of resource allocation and trajectory, we divide the problem $\text{P}_1$ into two sub-problems and solve them iteratively to achieve a sub-optimal solution using the alternating optimization.
Specifically, sub-problem 1 aims to optimize the subchannel and power allocation.
On the other hand, sub-problem 2 aims to optimize the UAV trajectory and velocity.


\begin{figure*}[t!]
\begin{align}\label{lag}
  & {{R}_{k}}[n]\ge  R_{k}^{i}[n]\triangleq\sum\limits_{j}^{{}}{\rho _{k,j}^{\mathrm{lic}}[n][{{\log }_{2}}(1+\frac{{{\beta }_{0}}p_{k,j}^{\mathrm{lic}}[n]}{X({{\left\| {{\mathbf{q}}_{i}}[n]-{{\mathbf{w}}_{k}} \right\|}^{2}}+{{H}^{2}})}})+\frac{-2{{\beta }_{0}}{{({{\mathbf{q}}_{i}}[n]-{{\mathbf{w}}_{k}})}^{H}}p_{k,j}^{\mathrm{lic}}[n]}{\ln 2(X({{\left\| {{\mathbf{q}}_{i}}[n]-{{\mathbf{w}}_{k}} \right\|}^{2}}+{{H}^{2}})+{{\beta }_{0}}p_{k,j}^{\mathrm{lic}}[n])}\frac{\mathbf{q}[n]-{{\mathbf{q}}_{i}}[n]}{({{\left\| {{\mathbf{q}}_{i}}[n]-{{\mathbf{w}}_{k}} \right\|}^{2}}+{{H}^{2}})}] \nonumber \\
 & +\sum\limits_{m}^{{}}{\rho _{k,m}^{\mathrm{unlic}}[n][{{\log }_{2}}(1+\frac{{{\beta }_{0}}p_{k,m}^{\mathrm{unlic}}[n]}{Y({{\left\| {{\mathbf{q}}_{i}}[n]-{{\mathbf{w}}_{k}} \right\|}^{2}}+{{H}^{2}})})}+\frac{-2{{\beta }_{0}}{{({{\mathbf{q}}_{i}}[n]-{{\mathbf{w}}_{k}})}^{H}}p_{k,m}^{\mathrm{unlic}}[n]}{\ln 2(Y({{\left\| {{\mathbf{q}}_{i}}[n]-{{\mathbf{w}}_{k}} \right\|}^{2}}+{{H}^{2}})+{{\beta }_{0}}p_{k,m}^{\mathrm{unlic}}[n])}\frac{\mathbf{q}[n]-{{\mathbf{q}}_{i}}[n]}{({{\left\| {{\mathbf{q}}_{i}}[n]-{{\mathbf{w}}_{k}} \right\|}^{2}}+{{H}^{2}})}]
\end{align}
\hrulefill \vspace*{1pt}
\end{figure*}

\vspace{-10pt}
\subsection{Spectrum Allocation and Power Control}
Given the fixed UAV trajectory, we first decompose problem $\textbf{P}_1$ into two sub-problems: (1) licensed subchannel allocation and power control, and (2) unlicensed subchannel allocation and power control.

\subsubsection{Licensed subchannel allocation and power control}
First, the licensed subchannel allocation and power control is given as
\begin{subequations}
\begin{align}
   \textbf{P}_{2}: \underset{\mathcal{A},\mathcal{P}}{\mathop{\max }}\,&\frac{1}{N}\sum\limits_{n=1}^{N}{\sum\limits_{k=1}^{K}\sum\limits_{j=1}^{J}{{{R}_{k,j}^{\mathrm{lic}}}[n]}} \\
  \mathrm{s.t.} \,  &\mathrm{C2}, \mathrm{C4}, \mathrm{C5}, \mathrm{C7}, \mathrm{C9},
\end{align}
\end{subequations}

Given the licensed subchannel allocation $\mathcal{A}$ and the licensed power control $\mathcal{P}$, the joint unlicensed spectrum allocation and power control can be represented as
\begin{subequations}
\begin{align}
   \textbf{P}_{3}: \underset{\mathcal{B},\mathcal{U}}{\mathop{\max }}\,&\frac{1}{N}\sum\limits_{n=1}^{N}{\sum\limits_{k=1}^{K}\sum\limits_{m=1}^{M}{{{R}_{k,m}^{\mathrm{unlic}}}[n]}} \\
  \mathrm{s.t.} \,  &\mathrm{C1}, \mathrm{C3}, \mathrm{C6}, \mathrm{C8}, \mathrm{C10}.
\end{align}
\end{subequations}
The solution of $\textbf{P}_2$ can be used to solve the sub-problem $\textbf{P}_3$, and vice versa repeatedly until converge.
We first study the solution of sub-problem $\textbf{P}_2$.

To tackle the problem $\textbf{P}_2$, we introduce auxiliary variable $\tilde{p}_{k,j}^{\mathrm{lic}}[n]=\rho _{k,j}^{\mathrm{lic}}[n]p_{k,j}^{\mathrm{lic}}[n],\forall k,j,n$. The equivalent problem
can be given as
\begin{subequations}
\begin{align}
   \textbf{P}_{2.1}: \underset{\mathcal{A},\mathcal{\tilde{P}}}{\mathop{\max }}\,&\frac{1}{N}\sum\limits_{n=1}^{N}{\sum\limits_{k=1}^{K}\sum\limits_{j=1}^{J}{{{\tilde{R}}_{k,j}^{\mathrm{lic}}}[n]}} \\
  \mathrm{s.t.} \,  &\mathrm{C4}, \mathrm{C5}, \mathrm{C7},\\
   &\widetilde{\mathrm{C2}}: \frac{1}{N} \sum_n \sum_k\left|h_j^{{\mathrm{up}}}[n]\right|^2 \tilde{p}_{k, j}^{ {\mathrm{lic}}}[n]<\Gamma_j^{{\mathrm{lic}}}, \forall j, \\
   & \mathrm{\widetilde{\mathrm{C9}}}: \sum_k \sum_j  \tilde{p}_{k, j}^{{\mathrm{lic}}}[n] \leq P_{\max }^{{\mathrm{lic}}}, \forall n,
\end{align}
\end{subequations}
where
$\tilde{R}_{k,j}^{{\mathrm{lic}}}[n]={{{\rho }_{k,j}}[n]{{\log }_{2}}(1+\frac{|h_{k}^{\mathrm{us}}[n]{{|}^{2}}\tilde{p}_{k,j}^{\mathrm{lic}}[n]}{(|h_{k}^{\mathrm{ps}}[n]{{|}^{2}}p_{j}^{\mathrm{P}}[n]+J_{k}+{{\sigma }_k^{2})\rho_{k,j}^{\mathrm{lic}}[n]}})}$ and $\tilde{\mathcal{P}}=\{\tilde{p}_{k,j}^{\mathrm{lic}}[n],\forall k,j,n\}$.
The main obstacle in solving $\textbf{P}_{2.1}$ arises from the binary licensed subchannel allocation $\mathrm{C1}$.
Therefore, to handle the binary constraint, we follow the approach as in \cite{chenlong} and relax the variable $\rho_{k,j}^{\mathrm{lic}}[n]$ such that it is a real
value between 0 and 1, which is given as
$0 \leq \rho_{k,j}^{\mathrm{lic}}[n] \leq 1, \forall k, j, n.$
In the following, we derive the Lagrangian function of  $\textbf{P}_{2.1}$, given as
\begin{align}
  & \mathcal{L}(\boldsymbol{\omega}, \boldsymbol{\theta}, \mathcal{A},\mathcal{\tilde{P}}) \nonumber \\
  &=\frac{1}{N}\sum\limits_{k}^{{}}\sum\limits_{j}^{{}}\sum\limits_{n}^{{}}\tilde{R}_{k,j}^{\mathrm{lic}}[n] +\sum\limits_{n}^{{}}{{{\theta }_{n}}(P_{\max }^{\mathrm{lic}}-\sum\limits_{k}^{{}}{\sum\limits_{j}^{{}}{\tilde{p}_{k,j}^{\mathrm{lic}}[n]}})}\nonumber \\
  &+\sum\limits_{j}^{{}}{{{\omega }_{j}}(\Gamma _{j}^{\mathrm{lic}}-\frac{1}{N}\sum\limits_{n}^{{}}{\sum\limits_{k}^{{}}{{{\left| h_{j}^{\mathrm{up}}[n] \right|}^{2}}\tilde{p}_{k,j}^{\mathrm{lic}}[n]}}}),
\end{align}
where $\boldsymbol{\omega}=\{\omega_{j},  \forall j\}$ and $\boldsymbol{\theta}=\{\theta_{n}, \forall n\}$ denote the Lagrange
multipliers for constraints $\mathrm{\widetilde{\mathrm{C2}}}$ and $\mathrm{\widetilde{\mathrm{C9}}}$, respectively.
Constraints $\mathrm{C4}$, $\mathrm{C5}$, and $\mathrm{C7}$ will be considered in the Karush-Kuhn-Tucker (KKT) conditions when deriving the optimal solution in the following \cite{rui_zhang}.
Then, the dual problem is given by
\begin{align}
\mathcal{D}=\underset{ \boldsymbol{\omega}, \boldsymbol{\theta} \geq 0}{\operatorname{minimize}} \  \underset{\mathcal{A}, \tilde{\mathcal{P}}} {\operatorname{maximize}} \mathcal{L}(\boldsymbol{\omega}, \boldsymbol{\theta}, \mathcal{A},\mathcal{\tilde{P}}).
\end{align}

Subsequently, the dual problem is solved iteratively via dual decomposition. In particular, the dual problem is decomposed into two nested layers:
Layer 1, maximizing the Lagrangian over the licensed subchannel allocation $\mathcal{A}$ and power allocation $\tilde{\mathcal{P}}$, Layer 2,
minimizing the Lagrangian function over $\omega$ and $\theta$ for a given $\mathcal{A}$ and $\tilde{\mathcal{P}}$.

\textit{Solution of Layer 1 (Licensed Subchannel and Power Allocation)}:
$p_{k,j}^{*,\mathrm{lic}}[n]$ and $\rho_{k,j}^{*,\mathrm{lic}}[n]$ denote the optimal solutions of sub-problem 1.
Then, the optimal power allocation for SU $k$ on subchannel $j$ at time slot $n$ is given by
\begin{align}\label{lic_p_opt}
p_{k,j}^{*,\mathrm{lic}}[n]=[\frac{1}{{\ln 2({\omega _j}{{\left| {h_j^{\mathrm{up}}[n]} \right|}^2} + {\theta _n})}} - \frac{{|h_{k}^{\mathrm{ps}}[n]{|^2}p_j^{P}[n] + {J_k} + {\sigma ^2}}}{{|h_k^{\mathrm{us}}[n]{|^2}}}]^+,
\end{align}
where $J_k={\left| {{{\bf{g}}_k^H}[n]{{\bf{w}}_k}[n]} \right|^2}$.
For a given $p_{k,j}^{*,\mathrm{lic}}[n]$, the sub-problem is combinatorial in the variable, where the Hungarian method \cite{shermen} is used to obtain
the optimal licensed subchannel allocation $\rho_{k,j}^{*,\mathrm{lic}}[n]$.

\textit{Solution of Layer 2 (Master Problem)}:
To solve Layer 2 master minimization problem, the gradient method is adopted and the Lagrange multipliers can be updated by
\begin{subequations}
\begin{align}\label{omege_update}
  & {{\omega }_{j}}(t)={{\omega }_{j}}(t)-\alpha_{1}(t) (\Gamma _{j}^{\mathrm{lic}}-\frac{1}{N}\sum\limits_{n}^{{}}{\sum\limits_{k}^{{}}{{{\left| h_{j}^{\mathrm{us}}[n] \right|}^{2}}\tilde{p}_{k,j}^{\mathrm{lic}}[n]}}), \\ \label{theta_update}
 & {{\theta }_{n}}(t)={{\theta }_{n}}(t)-\alpha_{2}(t) (P_{\max }^{\mathrm{lic}}-\sum\limits_{k}^{{}}{\sum\limits_{j}^{{}}{\tilde{p}_{k,j}^{\mathrm{lic}}[n]}}),
\end{align}
\end{subequations}
where $t \geq 0$ is the iteration index for sub-problem 1 and $\alpha_{u}(t), u=\{1,2\}$ are step sizes satisfying the infinite
travel condition [20].
Then, the updated Lagrangian multipliers are used for solving the Layer 1 sub-problem via updating the resource allocation policies.

\subsubsection{Unlicensed spectrum allocation and power control}
Similarly, in order to solve sub-problem $\textbf{P}_{3}$, the auxiliary variable $\tilde{p}_{k,j}^{\mathrm{unlic}}[n]=\rho _{k,m}^{\mathrm{unlic}}[n]p_{k,m}^{\mathrm{unlic}}[n],\forall k,m,n$ is introduced and subchannel allocation variable $\rho_{k,m}^{\mathrm{unlic}}[n]$ is relaxed, given as
$0 \leq \rho_{k,m}^{\mathrm{unlic}}[n] \leq 1, \forall k, m, n.$
Then, the sub-problem $\textbf{P}_{3}$ can be transformed given by
\allowdisplaybreaks
\begin{subequations}
\begin{align}
   \textbf{P}_{3.1}&: \underset{\mathcal{B},\tilde{\mathcal{U}}}{\mathop{\max }}\,\sum\limits_{n=1}^{N}{\sum\limits_{k=1}^{K}\sum\limits_{m=1}^{M}{{{\tilde{R}}_{k,m}^{\mathrm{unlic}}}[n]}} \\
  \mathrm{s.t.} \,  &\mathrm{C6},\\
  &\mathrm{\widetilde{C1}}: \frac{1}{N}\sum\limits_n^{} { {R_{k}^{\mathrm{unlic}}[n]} }  \ge {[R_k^{\min } - \frac{1}{N}\sum\limits_n^{} {{R_{k}^{\mathrm{lic}}[n]} } ]^ + },\forall k,\\
  &\widetilde{\mathrm{C3}}: \frac{1}{N} \sum_n \sum_k \left|h_m^{\mathrm{uw}}[n]\right|^2 \tilde{p}_{k, m}^{\mathrm{unlic}}[n]<\Gamma_m^{ {\mathrm{unlic}}}, \forall m, \\
& \mathrm{\widetilde{C8}}: 0 \leq \rho_{k,m}^{\mathrm{unlic}}[n] \leq 1, \forall k, \forall m, \forall n,\\
& \widetilde{\mathrm{C10}}: \sum_k \sum_m \tilde{p}_{k, m}^{{\mathrm{unlic}}}[n] \leq P_{\max }^{{\mathrm{unlic}}}, \forall n,
\end{align}
\end{subequations}

Note that the problem in $\textbf{P}_{3.1}$ is jointly convex with respect to all optimization variables and the Slater's condition is satisfied [37].
Therefore, strong duality holds, i.e., the gap between the optimal value and that of its dual problem is zero [37].
Therefore, the Lagrangian dual method is adopted to achieve the optimal unlicensed subchannel allocation.
Besides, to find the optimal subcarrier allocation, we take the derivative of the Lagrangian function w.r.t. $\rho_{k,m}^{\mathrm{lic}}[n]$
which yields $M_{k,m}[n]$. Due to the constraint $\mathrm{C6}$, the optimal unlicensed subchannel allocation is given by
\begin{align}\label{rho_unlic}
\rho_{k,m}^{*,\mathrm{unlic}}[n]=\left\{\begin{array}{ll}1, & k^*=\underset k \max \left(M_{k,m}[n]\right), \\ 0, & \text{otherwise},\end{array} \quad \forall m, n.\right.
\end{align}

\subsection{UAV trajectory optimization}
The UAV trajectory is optimized given the subchannel and power allocation.
The trajectory of the UAV can be optimized by solving the following problem
\begin{subequations}
\begin{align}
   \textbf{P}_{4}: &\underset{\mathcal{Q}}{\mathop{\max }}\,\sum\limits_{n=1}^{N}{\sum\limits_{k=1}^{K}{{{R}_{k}^{}}[n]}} \\
  \mathrm{s.t.} \,  &\mathrm{C1}, \mathrm{C2},\mathrm{C3},\mathrm{C11},
\end{align}
\end{subequations}













It is noted that problem $\textbf{P}_{4}$ is a non-convex problem since the objective function is a
non-concave function and ${\mathrm{C2}}$ and ${\mathrm{C3}}$ are non-convex constraints.
To make $\textbf{P}_{4}$  more tractable, the slack variables $\xi _{j}^{\mathrm{cell}}[n]$ and $\xi _{m}^{\mathrm{Wifi}}[n]$ are firstly introduced, and constraint $\mathrm{C2}$ and $\mathrm{C3}$ can be rewritten as
\begin{subequations}
\begin{align}
 &\widetilde{\mathrm{C2}}: \frac{1}{N}\sum\limits_n^{} {\sum\limits_k^{} {\frac{{{\rho _{k,j}^{\mathrm{lic}}}[n]{\beta _0}p_{k,j}^{\mathrm{lic}}[n]}}{{{H^2} + \xi _j^{\mathrm{cell}}[n]}}} }  < \Gamma _j^{\mathrm{lic}},\forall j,\\
 &\widetilde{\mathrm{C3}}: \frac{1}{N}\sum\limits_n^{} {\sum\limits_k^{} {\frac{{{\rho _{k,m}^{\mathrm{unlic}}}[n]{\beta _0}p_{k,m}^{\mathrm{unlic}}}}{{{H^2} + \xi _m^{\mathrm{Wifi}}[n]}}} [n]}  < \Gamma _m^{\mathrm{unlic}},\forall m,
 \end{align}
\end{subequations}
where the wireless channel between the C-UAV and the ground users is dominated by the LoS link, $\beta_0$ represent the channel
power gain at the reference distance of 1 meter \cite{chenlong}.
Then, problem $\textbf{P}_4$ can be rewritten as
\begin{subequations}
\begin{align}
   \textbf{P}_{4.1}:& \underset{\mathcal{Q}, \xi _{j}^{\mathrm{cell}}[n], \xi _{j}^{\mathrm{Wifi}}[n]}{\mathop{\max }}\,\sum\limits_{n=1}^{N}{\sum\limits_{k=1}^{K}{{{R}_{k}}[n]}} \\
  \mathrm{s.t.} \,  &{\mathrm{C1}}, \widetilde{\mathrm{C2}}, \widetilde{\mathrm{C3}}, {\mathrm{C11}},\\
& \mathrm{C12}: \xi _j^{\mathrm{cell}}[n] \le {\left\| {{\bf{q}}[n] - {\bf{w}}_j^{\mathrm{cell}}} \right\|^2},\forall n,\\
& \mathrm{C13}: \xi _m^{\mathrm{Wifi}}[n] \le {\left\| {{\bf{q}}[n] - {\bf{w}}_m^{\mathrm{Wifi}}} \right\|^2},\forall n.
\end{align}
\end{subequations}

It is noted that $\mathrm{C12}$ and $\mathrm{C13}$ are non-convex constraints, and the objective function is non-concave.
Thus, problem $\textbf{P}_{4.1}$ is still a non-convex optimization problem.
To solve the problem, we propose the successive convex optimization based algorithm to obtain an approximate solution.
Specifically, at each iteration, problem $\textbf{P}_{4.1}$ is approximated by a solvable function at a given point $\mathbf{q}_{i}[n]$, where $i$ is the number of iterations.
Therefore, $R_{k}^{}[n]$ is approximated as the lower bound $R_{k}^{i}[n]$, as shown in (\ref{lag}), where $X = |h_{k}^{\mathrm{ps}}[n]{|^2}p_j^{P}[n] + {J_k} + {\sigma ^2}$ and $Y = |h_{k}^{\mathrm{ws}}[n]{|^2}p_m^{\mathrm{Wifi}}[n] + {J_k} + {\sigma ^2}$.

The constraints $\mathrm{C12}$ and $\mathrm{C13}$ are approximated as a convex set by using the first-order Taylor expansion at the given local point in the $i$th iteration, given as
\begin{subequations}
\begin{align}
{{\left\| \mathbf{q}[n]-\mathbf{w}_{j}^{\mathrm{cell}} \right\|}^{2}}&\ge {{\left\| {{\mathbf{q}}_{i}}[n]-\mathbf{w}_{j}^{\mathrm{cell}} \right\|}^{2}} \nonumber \\
&+2{{({{\mathbf{q}}_{i}}[n]-\mathbf{w}_{j}^{\mathrm{cell}})}^{H}}(\mathbf{q}[n]-{{\mathbf{q}}_{i}}[n]),\\
{{\left\| \mathbf{q}[n]-\mathbf{w}_{m}^{\mathrm{Wifi}} \right\|}^{2}}&\ge {{\left\| {{\mathbf{q}}_{i}}[n]-\mathbf{w}_{m}^{\mathrm{Wifi}} \right\|}^{2}} \nonumber \\
&+2{{({{\mathbf{q}}_{i}}[n]-\mathbf{w}_{m}^{\mathrm{Wifi}})}^{H}}(\mathbf{q}[n]-{{\mathbf{q}}_{i}}[n]).
\end{align}
\end{subequations}
Therefore, problem $\textbf{P}_{4.1}$ can be approximated given as
\begin{subequations}
\begin{align}
   \textbf{P}_{4.2}: &\underset{\mathcal{Q}, \xi _{j}^{\mathrm{cell}}[n], \xi _{j}^{\mathrm{Wifi}}[n]}{\mathop{\max }}\,\sum\limits_{n=1}^{N}{\sum\limits_{k=1}^{K}{{{R}_{k}^{i}}[n]}} \\
  \mathrm{s.t.} \,  &\bar{\mathrm{C1}}: \frac{1}{N} \sum_n R_k^{i}[n] \geq R_k^{\min }, \forall k, \\
 & \widetilde{\mathrm{C2}}, \widetilde{\mathrm{C3}}, \mathrm{C11},\\
& \widetilde{\mathrm{C12}}: \xi _j^{\mathrm{cell}}[n] \le {\left\| {{{\bf{q}}_i}[n] - {\bf{w}}_j^{\mathrm{cell}}} \right\|^2} \nonumber \\
& \quad + 2{({{\bf{q}}_i}[n] - {\bf{w}}_j^{\mathrm{cell}})^H}({\bf{q}}[n] - {{\bf{q}}_i}[n]), \forall n, \forall j, \\
& \widetilde{\mathrm{C13}}: \xi _m^{\mathrm{Wifi}}[n] \le {\left\| {{{\bf{q}}_i}[n] - {\bf{w}}_j^{\mathrm{Wifi}}} \right\|^2} \nonumber \\
& \quad + 2{({{\bf{q}}_i}[n] - {\bf{w}}_j^{\mathrm{Wifi }})^H}({\bf{q}}[n] - {{\bf{q}}_i}[n]), \forall n, \forall m.
\end{align}
\end{subequations}

Since $\textbf{P}_{4.2}$ is a concave
function with respect to $\mathcal{Q}$, $\mathrm{C1}$ is a convex constraint, $\widetilde{\mathrm{C2}}$ and $\widetilde{\mathrm{C}3}$ are convex fractional constraints, $\widetilde{\mathrm{C12}}$ and $\widetilde{\mathrm{C13}}$ are linear constraints, $\mathrm{C11}$ is a convex quadratic constraint, problem $\textbf{P}_{4.2}$ is a convex optimization problem,
which can be solved efficiently by using CVX.

\begin{figure*}
\centering
\subfigure[ ]{
\includegraphics[height=1.45in]{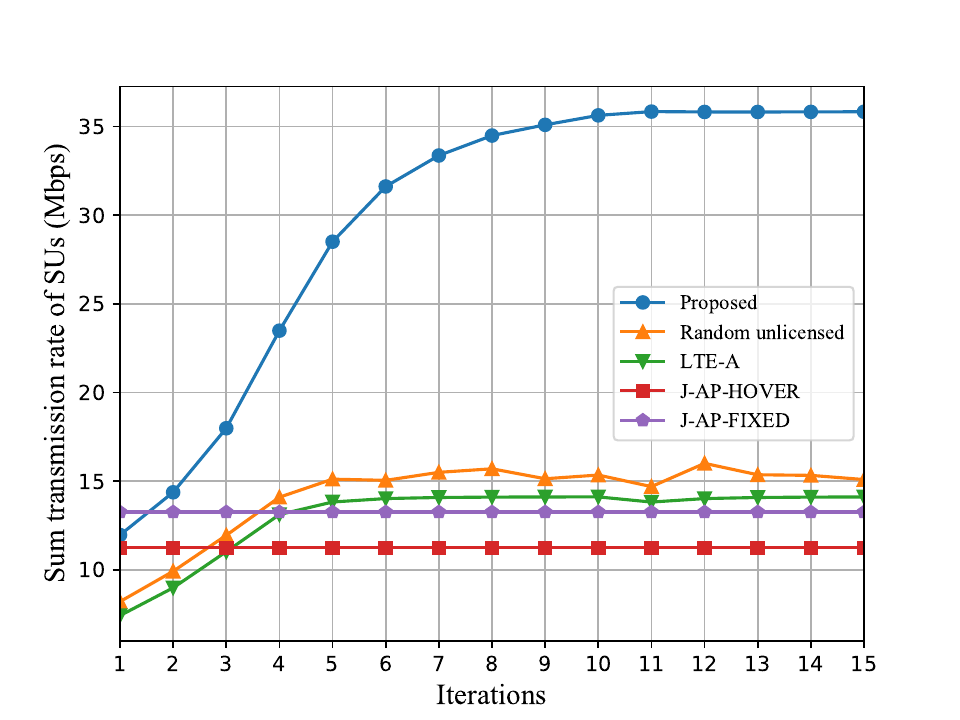}
\hspace{5mm}\label{iter}
}
\subfigure[ ]{
\includegraphics[height=1.45in]{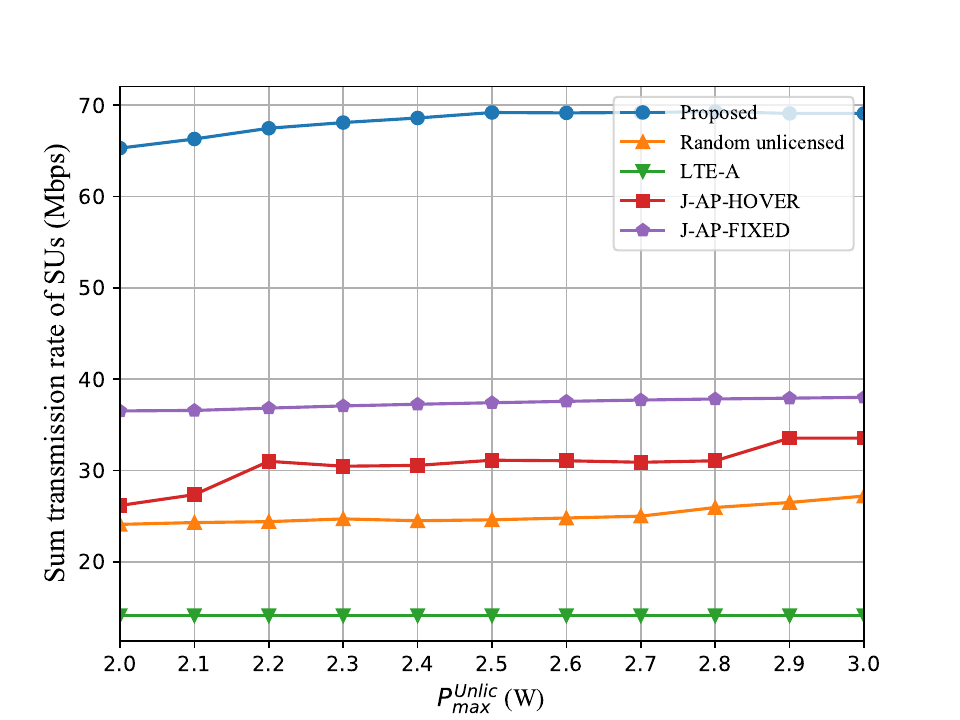}
\hspace{5mm} \label{p_vs_rate}
}
\subfigure[ ]{
\includegraphics[height=1.45in]{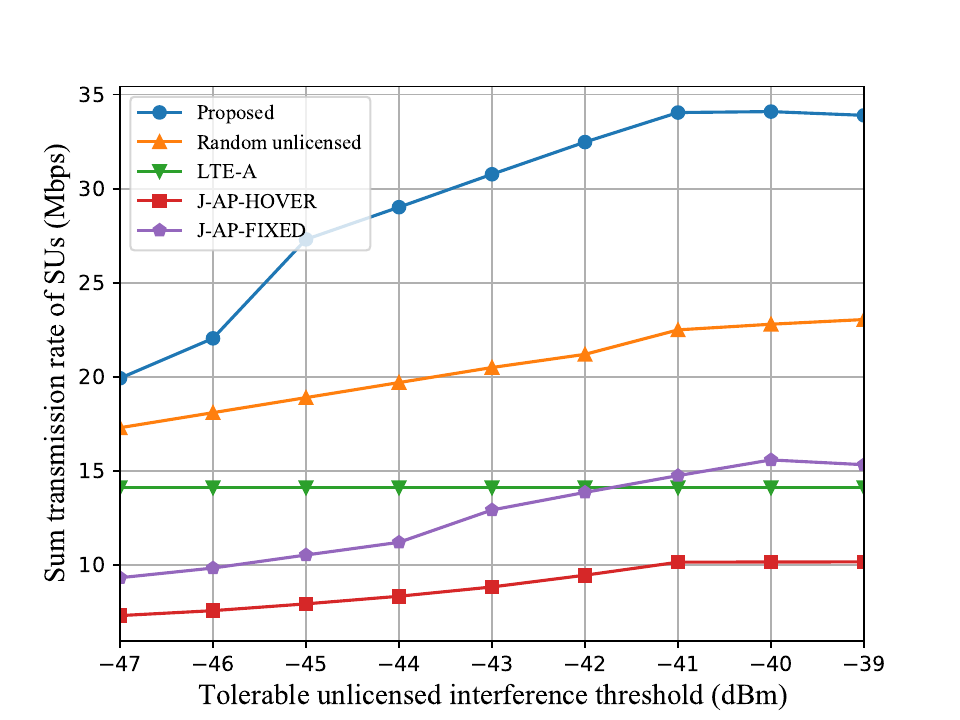}
\hspace{5mm}\label{threshold_vs_rate}
}
\DeclareGraphicsExtensions.
\vspace{-12pt}
\caption{(a) Sum transmission rate versus the number of iterations with the maximum transmit power of unlicensed spectrum is $P_{\mathrm{max}}^{\mathrm{unlic}} = 27.6$ dBm and the unlicensed interference threshold is $\Gamma_m^{ \mathrm{unlic}}=-35$ dBm.
(b) Sum transmission rate versus maximum transmit power $P_{\mathrm{max}}^{\mathrm{unlic}}$ for different schemes. The maximum tolerable interference for unlicensed Wi-Fi network is $\Gamma_m^{ \mathrm{unlic}}=-35$ dBm. (c) Sum transmission rate versus tolerable unlicensed interference threshold $\Gamma_m^{ \mathrm{unlic}}$ for different schemes.}
\end{figure*}

\vspace{-1pt}

\section{Simulation Results} \label{Simulation Results}
In this section, simulation results are presented to evaluate the performance of our proposed algorithm.
The simulation settings are based on the work in \cite{shermen} and \cite{chenlong}.
For comparison, four benchmark schemes are considered.
The first benchmark scheme is assumed that the subchannel allocation of the unlicensed spectrum are generated randomly, which is
denoted by Random unlicensed \cite{chenlong}.
The second one is the classical LTE-A scheme, where the C-UAV serves the SUs in the licensed spectrum only \cite{shermen}.
For the third benchmark scheme, the UAV statically hovers over the whole service period, which is denoted by J-AP-HOVER  \cite{iot}.
The forth benchmark is marked as J-AP-FIXED, where the C-UAV trajectory is designed based on fixed random trajectory \cite{iot}.

Fig. \ref{iter} illustrates the convergence behavior of the proposed alternating optimization algorithm  for the maximization of the secondary network sum transmission rate.
It is seen that the proposed algorithm can converge to a stationary point within 8 iterations.
Despite the benchmark schemes achieve faster convergence, the sum transmission rate achieved by the proposed algorithm is superior than that of the benchmark schemes, which verified the effectiveness of our proposed algorithm.
Moreover, since the UAV trajectory remains fixed in both J-AP-HOVER and J-AP-FIXED, there is no need for alternate optimization.
Therefore, the sum transmission rate remains constant across iterations.

Fig. \ref{p_vs_rate} shows the sum transmission rate of all five schemes versus the maximum unlicensed transmit power from 1.98 W to 3.08 W.
It can be seen that the sum rate of SUs increases with the maximum transmit power.
Moreover, the proposed scheme outperforms all the benchmark schemes in terms of the achievable sum transmission rate.
Furthermore, the interference caused to the unlicensed Wi-Fi network exceeds the tolerable threshold when the transmit power is 2.5 W. Therefore, the sum transmission rate gain in the secondary network is limited.
It demonstrates that the proposed algorithm can efficiently improve the secondary network sum transmission rate while protecting the unlicensed Wi-Fi network.

Fig. \ref{threshold_vs_rate} illustrates the secondary network sum transmission rate achieved with
those five schemes versus the tolerable unlicensed interference threshold.
Firstly, it can be seen that the transmission rate increases with the interference power threshold since a higher power can be utilized to
improve the communication performance.
In fact, the subchannel and power allocation of the C-UAV are restricted significantly when the interference power threshold is low.
Then, the transmission rate remain constant when $\Gamma_m^{ \mathrm{unlic}}$ is large enough.
The reason is that the transmit power is dominantly limited by the peak transmit power constraints, thus further increasing $\Gamma_m^{ \mathrm{unlic}}$ can not enlarge the feasible region.

\section{Conclusion} \label{Conclusion}
In this paper, the sum-rate maximization problem in joint licensed and unlicensed spectrum sharing UAV network was studied.
The transmit power, subchannel allocation, and UAV trajectory were jointly optimized to maximize the secondary network sum transmission rate.
To tackle the challenging formulated non-convex optimization problem, an efficient iterative algorithm was proposed.
Simulation results demonstrated the efficiency of our proposed algorithm.
It was also shown that our proposed algorithm can achieve higher sum-rate compared with the benchmark schemes.



\end{document}